\documentclass[preprint,1p]{elsarticle}
\usepackage{amssymb,graphicx}

\usepackage{bm}
\usepackage{graphics}
\usepackage{mathtools}
\usepackage{amsmath,amsbsy,amssymb,amsfonts}  
\usepackage[parfill]{parskip}  

\newcommand*{\half}{\frac{1}{2}}

\newcommand*{\eff}{\text{eff}}

\DeclareMathOperator{\Tr}{Tr}

\newcommand*{\ket}[1]{| #1 \rangle}

\newcommand*{\avg}[1]{\langle #1 \rangle}

\biboptions{sort&compress}

\begin{document}
\begin{frontmatter}

\title{Absorption-dispersion in a three-level electromagnetically induced
transparency medium including near dipole-dipole interaction effects}

\author{Amitabh Joshi\corref{cor1}}
\ead{mcbamji@gmail.com}
\address{Department of Physics and Optical Engineering, Rose-Hulman Institute
of
Technology, Terre Haute, Indiana 47803, USA}

\author{Juan D. Serna\corref{cor2}}
\ead{juan.serna@scranton.edu}
\address{Department of Physics and Electrical Engineering, University of
Scranton, Scranton, Pennsylvania 18510, USA}

\cortext[cor1]{Principal corresponding author}
\cortext[cor2]{Corresponding author}

\begin{abstract}
Dynamical evolution and electromagnetically induced transparency (EIT) is
investigated here in a three-level $\lambda$-type atomic system including
near-dipole-dipole interaction among atoms. The system is driven by the probe
and coupling fields. Exact numerical solutions under steady-state condition are
given for the density operator equation to get information about population in
various levels and the linear susceptibility of probe-transition. Also, obtained
are the closed form expressions for linear and third order non-linear
susceptibilities for the probe transition under perturbation approximation.
\end{abstract}

\end{frontmatter}

\section{\label{sec:intro}Introduction}
It has been shown in the late eighties that the propagation of an
electromagnetic field with a medium composed of two-level atoms can generate
near dipole-dipole (NDD) interaction. Such NDD effects can result in the
inversion-dependent-chirping of the single atom resonance frequency of such
two-level atomic dipole system. The NDD interactions give rise to a local effect
that modifies the microscopic field coupling the atom and which is obtained from
the macroscopic field and the induced polarization~\cite{Bowden:1993}. The
significant contribution of the NDD interaction comes from the entities enclosed
in a tiny volume of the order of a cubic wavelength, and that is prominent in a
dense medium. The use of the modified Maxwell-Bloch equation allowed to predict
many interesting results caused by NDD effects. For example, invariant pulse
propagation that departs from the hyperbolic secant pulse shape (with pulse area
different from $2\pi$) related to self-induced transparency
(SIT)~\cite{Bowden:91} and self-phase modulation in SIT~\cite{Stroud:67}. Other
relevant results include the observation of intrinsic optical bistability (IOB)
when the atomic number density and the oscillator strengths are very
high~\cite{Ben-Aryeh:34}; enhancement of gain in systems showing inversionless
lasing; optical switching; among others. When a sample of atoms interacts with
the external driving field, then the generated reaction field due to the induced
dipoles in this samples works against the applied field leading to a decrease in
the net field. If the external driving field is stronger than the generated
reaction field due to the dipole-dipole interaction, then the manifestation of
the suppression of reaction field can be observed as a first-order phase
transition far away from the equilibrium
condition~\cite{Inguva:41,Ben-Aryeh:61}.

Modified nonlinear Maxwell-Bloch equations are required to describe the
interaction of the propagating electromagnetic field in a dense two-level
medium~\cite{Bowden:1993}. In an optically dense medium, the near dipole-dipole
interaction among atoms (occurring at microscopic scale) plays a significant
role and leads to the renormalization of the resonance frequency of transition.
This renormalization is governed by the population inversion in the two-level
system. However, to get a relationship between the macroscopic electric field
$\mathbf{E}_M$ and the polarization $\mathbf{P}$ to the microscopic field
$\mathbf{E}_m$ (causing the excitation in the atomic system through the dipole
interaction), the Lorentz-Lorentz relation
\begin{equation}\label{Eq:Lorentz-Lorentz}
\mathbf{E}_m = \mathbf{E}_M + \frac{4\pi}{3} \mathbf{P}
\end{equation}
is required. This equation is valid for homogeneous and isotropic media of the
static fields. According to the extinction theorem~\cite{Ewald:49}, such an
equation is also valid for the monochromatic time-dependent propagating field in
a linear, homogeneous, and isotropic medium. However, in an optically dense
medium where many interesting nonlinear effects can be studied, one needs to get
a proper relationship between the microscopic and the macroscopic fields.

Maxwell's equations along with Eq.~(\ref{Eq:Lorentz-Lorentz}) (related to the
local field correction) provide a Clausius-Mossotti relation between the
microscopic polarizability $\beta$, and the macroscopic dielectric parameter
$\epsilon_M$, of any solid, liquid or gaseous medium. This relation takes the
form~\cite{Bowden:1993}
\begin{equation}\label{Eq:Clausius-Mossotti}
\beta = \frac{3}{4\pi N} \left(\frac{\epsilon_M-1}{\epsilon_M+2}\right),
\end{equation}
where $N$ is the number of entities (atoms or molecules) per unit volume.
Causality and retardation phenomena are essential to explain the propagation of
time-dependent fields. In the literature, it has been shown using the extinction
theorem that for a linear, homogeneous, and isotropic medium,
Eq.~(\ref{Eq:Lorentz-Lorentz}) works well and hence
Eq.~(\ref{Eq:Clausius-Mossotti}) is also appropriate for such
medium~\cite{Bowden:1993}. The propagation of a field in a dense and nonlinear
medium consisting of multi-level atoms requires modified atomic and field
equations to include the effect of induced dipole-dipole interaction in these
equations.

The Maxwell's wave equation provides the relationship between the macroscopic
electric field $\mathbf{E}_M$, and the macroscopic polarization $\mathbf{P}$
\begin{equation}
\nabla^2 \mathbf{E}_M -
\frac{1}{c^2}\frac{\partial^2\mathbf{E}_M}{\partial t^2} =
\frac{4\pi}{c^2}\frac{\partial^2\mathbf{P}}{\partial t^2},
\end{equation}
with $c$ the speed of light in vacuum. The vector quantities
$\mathbf{E}_M$ and $\mathbf{P}$ are waves traveling in the $z$-direction and
expressed as
\begin{equation}
\begin{split}
\bf{E}_M &= \varepsilon\,e^{-i(\omega t-k_z z)} + c.c.\\
\bf{P} &= \wp\,e^{-i(\omega t-k_z z)} + c.c.
\end{split}
\end{equation}
with wave vector $k_z$, frequency $\omega$, and slowly varying quantities
${\varepsilon}$ and ${\wp}=i\mu N D_{ab}$. $N$ is the density of two-level atoms
in the medium and $\mu$ is the transition dipole matrix element.\\

In general, the microscopic field interacting with the atomic dipole is not
identical with the macroscopic field appearing in Maxwell's equations. This
difference is because the field driving the atom does not contain the local
field of the atom. On the other hand, the macroscopic field of Maxwell's
equations does include the local field. Hence, it is essential to get a
relationship between the microscopic and macroscopic field when the atomic
system is optically dense (which means a large number of atoms within a cubic
resonance wavelength)~\cite{Bowden:1993}.

In previous works~\cite{Bowden:1993,Ben-Aryeh:61} for the two-level system, the
Maxwell-Bloch equations for the dense medium under the slowly-varying-envelope
approximation for the field were obtained for a homogeneous medium that
contained a large number of atoms within a small volume determined by the cube
of the resonance wavelength. In earlier work~\cite{Dowling:70}, the effects of
near dipole-dipole interaction on a three-level system undergoing lasing without
inversion were studied, and enhancements in inversionless gain and refractive
index without absorption were predicted. However, in that work, the authors
treated the problem differently from the way we intend to show in this present
work on a three-level system. In another work, the effect of dipole-dipole
interaction has been discussed in the cavity quantum electrodynamics of
two-level atoms~\cite{Joshi:91,Joshi:44}.

This paper is organized as follows: Section~\ref{sec:model} describes the model
under consideration followed by some analytical results for linear and
third-order susceptibilities of the probe transition in
Section~\ref{sec:analytical}. Numerical results for absorption and dispersion
are discussed in Section~\ref{sec:numerical} by solving the exact density matrix
equation. In Section~\ref{sec:conclusion} some concluding remarks are provided.

\section{\label{sec:model}The model}
\begin{figure}
  \centering
  \includegraphics[scale=1]{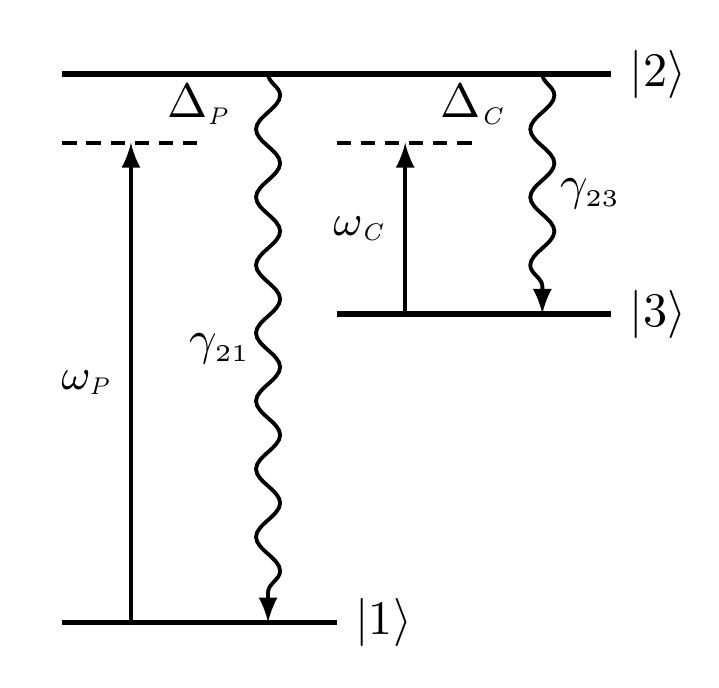}
  \caption{Diagram of the three-level system in a $\Lambda$-configuration and
  driven by a probe and coupling lasers of frequencies $\omega_P$ and
  $\omega_C$, respectively.}
  \label{Fig:1}
\end{figure}

In this work, we extend the development of atomic density operator equations for
a three-level atomic system when the fields are propagating in an optically
dense medium. Such a dense medium is characterized by a high number of atoms
within the volume determined by the cube of the resonance wavelength. The medium
consists of a three-level system in a $\lambda$-type configuration of its
levels~\cite{Joshi:45,Wang:49}. The model under consideration uses a
semiclassical approximation where the system interacts with the classical
electromagnetic fields of two lasers. The probe and coupling laser beams with
frequencies $\omega_{P}$ and $\omega_{C}$, respectively, interact with the
atomic transitions  $\omega_{21}$ and $\omega_{23}$, as shown in
Fig.~\ref{Fig:1}. The Liouville equations of density-matrix elements in the
dipole and rotating wave approximations are given by~\cite{Joshi:45,Wang:49}\
\begin{align}\label{Eq:Liouville}
\begin{split}
\dot{\rho_{22} - \rho_{11}} =&
-(\gamma_{23} + 2\gamma_{21})\rho_{22}
+ 2i\mu_{12}(\varepsilon^{P}_L)^*\rho_{21}
- 2i\mu_{12}(\varepsilon^{P}_L)\rho_{12} \\
& +\ i\mu_{23}(\varepsilon^{C}_L)^*\rho_{23}
- i\mu_{23}(\varepsilon^{C}_L)\rho_{32}
- \gamma_{31}(\rho_{33} - \rho_{11}), \\
\dot{\rho_{22} - \rho_{33}} =&
-(2\gamma_{23} + \gamma_{21})\rho_{22}
+ i\mu_{12}(\varepsilon^{P}_L)^*\rho_{21}
- i\mu_{12}(\varepsilon^{P}_L)\rho_{12} \\
& +\ 2i\mu_{23}(\varepsilon^{C}_L)^*\rho_{23}
- 2i\mu_{23}(\varepsilon^{C}_L)\rho_{32}
- \gamma_{31}(\rho_{11} - \rho_{33}), \\
\dot{\rho}_{23} =&
-(\gamma + i\Delta_{C})\rho_{23}
+ i\mu_{23}(\varepsilon^{C}_L)(\rho_{22} - \rho_{33})
- i\mu_{12}(\varepsilon^{P}_L)\rho_{13}, \\
\dot{\rho}_{21} =&
-(\gamma +i\Delta _{P})\rho_{21}
+ i\mu_{12}(\varepsilon^{P}_L)(\rho_{22} - \rho_{11})
- i\mu_{23}(\varepsilon^{C}_L)\rho_{31}, \\
\dot{\rho}_{31} =&
-\left[\gamma_{31} + i(\Delta_{P} -
\Delta_{C})\right]\rho_{31} - i\mu_{23}(\varepsilon^{C}_L)^*\rho_{21}
+ i\mu_{12}(\varepsilon^{P}_L)\rho_{32}.
\end{split}
\end{align}

In Eq.~(\ref{Eq:Liouville}), $\varepsilon_{L}^P$ and $\varepsilon _{L}^C$ are
complex, microscopic, slowly-varying electric field envelopes of the probe and
coupling fields, respectively. The radiative decay rates from levels $\ket{2}$
to $\ket{1}$ and $\ket{2}$ to $\ket{3}$ are $\gamma_{21}$ and $\gamma_{23}$,
respectively. The non-radiative decay rate between levels $\ket{3}$ and
$\ket{1}$ is $\gamma_{31}$. We also introduce $\gamma = \half(\gamma_{21}
+ \gamma_{32} + \gamma_{31})$. The Rabi frequencies  of the probe and coupling
fields are defined as $\Omega^{P} = 2\mu_{12}\varepsilon_{L}^P$ and $\Omega^{C}
= 2\mu_{32}\varepsilon_{L}^C$, respectively. The transition dipole matrix
elements for transitions between  levels $\ket{1}$ and $\ket{2}$ ($\ket{3}$) is
$\mu_{12}$ ($\mu_{23}$), which will be commonly represented by the symbol
$\mu_i$ (with $i=12, 23$) in the subsequent discussion.

Since we have two electromagnetic fields interacting with the three-level
system, we denote these microscopic fields by $\mathbf{E}_L^{i}$, with $i=P,C$,
and P and C denoting the probe and coupling field, respectively. We consider
fields to be linearly polarized and moving as plane waves. The three-level atom
is stationary and located at position $\mathbf{r}_l$. The microscopic field
is the sum of the external driving field and the reaction field of the induced
dipoles in the medium~\cite{Bowden:1993}
\begin{equation}\label{Eq:MicroscopicField_L}
\mathbf{E}_L^{i}(\mathbf{r}_l,t) =
\mathbf{E}_{ext}^{i}(\mathbf{r}_l,t) + \sum_{m=1}^{N}\aleph_{lm}^i
\exp\left[-i\mathbf{k}^i\cdot(\mathbf{r}_l - \mathbf{r}_m)\right]
\rho^i (t-r_{rl}/c), \quad (t=P,C).
\end{equation}
In Eq.~(\ref{Eq:MicroscopicField_L}), $\rho^P = \rho_{12}$ and $\rho^C =
\rho_{23}$, and the quantity $\aleph_{lm}^i$ is given by
\begin{equation}\label{Eq:Aleph}
\begin{split}
\aleph_{lm}^i = (3/2)\zeta^i \left\{\left[
\frac{\mathbf{p}_l^i}{|\mathbf{p}_l^i|}\cdot
\frac{\mathbf{p}_m^i}{|\mathbf{p}_m^i|} - \left(
\frac{\mathbf{p}_l^i}{|\mathbf{p}_l^i|}\cdot
\frac{\mathbf{r}_{lm}}{|\mathbf{r}_{lm}|}\right)\left(
\frac{\mathbf{p}_m^i}{|\mathbf{p}_m^i|}\cdot
\frac{\mathbf{r}_{lm}}{|\mathbf{r}_{lm}|}\right)\right]
Q_1^i(kr_{lm}) \right. \\
+ \left.\left(
\frac{\mathbf{p}_l^i}{|\mathbf{p}_l^i|}\cdot
\frac{\mathbf{r}_{lm}}{|\mathbf{r}_{lm}|}\right)\left(
\frac{\mathbf{p}_m^i}{|\mathbf{p}_m^i|}\cdot
\frac{\mathbf{r}_{lm}}{|\mathbf{r}_{lm}|}\right)
Q_2^i(kr_{lm}) \right\},
\end{split}
\end{equation}
with
\begin{align}\label{Eq:Q_i}
\begin{split}
Q_1^i(kr_{lm}) & = i e^{ik^ir_{lm}}\left[\frac{1}{(k^i)^3 r_{lm}^3}
- \frac{1}{k^ir_{lm}} + \frac{i}{(k^i)^2 r_{lm}^2}\right], \\
Q_2^i(k^i r_{lm}) & = 2i e^{-k^i r_{lm}}\left[-\frac{1}{(k^i)^3 r_{lm}^3}
+ i\frac{1}{(k^i)^2 r_{lm}^2}\right], \\
\zeta^i & =  \frac{2|\mu_i|^2(k^i)^3}{3\hbar};
\end{split}
\end{align}
In addition, $k^i = \omega^i/c$, $\mathbf{r}_{lm} = \mathbf{r}_l -
\mathbf{r}_m$, and $\mathbf{p}_l^i$ ($i=P,C$) are the dipole moments of the
probe and coupling transitions.

The microscopic field $\mathbf{E}_L^{i}$ has contributions of the nearby region
specified by the condition $r_{lm}\ll\lambda$, the intermediate region, when
$r_{lm}=\lambda$, and the far-off region, when $r_{lm}\gg\lambda$. The
near-field defined by the parameter $E^i_{ne}(\mathbf{r}_l,t)$ is due to the
atomic dipoles contained in a slab of thickness $\Delta z$ around the position
of the atom, situated at $r_l$. This field is a sort of microscopic contribution
so that the other fields can be clubbed as the macroscopic
field~\cite{Bowden:1993}. The field $E^i_{ne}(\mathbf{r}_l,t)$ gives non-zero
contribution under the plane wave propagation condition in a homogeneous
isotropic medium. In the following, we provide the estimates of these fields in
different regions for the three-level system under consideration, just for the
sake of completeness on the lines of Ref.~\cite{Bowden:1993}, where a two-level
medium was considered.

\subsection{Microscopic Field in Nearby region}
We estimate the microscopic fields by confining the atomic sample in a
cylindrical shape geometry with the cylindrical axis along the $z$-direction. To
make calculations simple, we assume the electromagnetic field as a plane wave
propagating along the positive $z$-direction and the atomic location $r_l$ on
the $z$-axis. We also consider the dipoles of the medium to be pointing in the
$x$-direction and plane polarized. Our goal is to estimate the field produced by
all such identical dipoles (within a cylinder of diameter $d$) at location
$r_l$. To evaluate the microscopic field, we replace the summation in
Eq.~(\ref{Eq:MicroscopicField_L}) by an integral under the continuum limit to
obtain~\cite{Bowden:1993}
\begin{equation}\label{Eq:MicroscopicField_ne}
E^i_{ne}(z,t) = \rho^i G^i(r, z, s) = (3/2)\rho^i\zeta^i N
\int_{0}^{2\pi}d\theta
\int_{-d/2}^{d/2} dz
\int_{s_m}^{s_M} s\,ds\,H^i,
\end{equation}
with
\begin{equation}\label{Eq:H_i}
H^i = Q^i_1(k^i r) \sin^2 \theta + Q^i_2(k^i r) \cos^2 \theta
+ \frac{|z-z_l|^2}{r^2} [Q^i_1(k^i r)- Q^i_2(k^i r)] \cos^2\theta,
\end{equation}
and $r = (s^2 + |z-z_l|^2)^{1/2}$. Here, $\rho^i$ (with $i=P,C$) represents the
density matrix elements $\rho_{21}$ and $\rho_{23}$; $s_m$ and $s_M$ are the
minimum and maximum values of $s$. We neglect the contribution of a small
volume about the atom and around $z = z_l$, the quantity $r_{lm}$ is approaching
to $r$ and $d \ll s_M$ while carrying out the integration and get
\begin{equation}\label{Eq:Gi_1}
\avg{G^i} = \frac{2i\pi N \zeta^i}{(k^i)^3} + \frac{3\pi N \zeta^i d}{(k^i)^2},
\quad \text{with} \quad i=P,C.
\end{equation}
Clearly, Eq.~(\ref{Eq:Gi_1}) is a complex quantity whose real part depends on
the diameter of the cylinder, $d$, but the imaginary part is independent of it.
After substituting the value of $\zeta^i$ in Eq.~(\ref{Eq:Gi_1}),
we get
\begin{equation}\label{Eq:Gi_2}
\avg{G^i} = \frac{4i\pi |\mu_i|^2 N}{3\hbar}
+ \frac{2\pi N |\mu_i|^2 d k^i}{\hbar} = \epsilon_i + 2\gamma^D_i.
\end{equation}
In the limiting condition, $2\pi d / \lambda^i = k^i \ll 1$, the
contribution from the second term in Eq.~(\ref{Eq:Gi_2}) is negligibly small
and thus the contribution from near dipoles coming under the condition $\Delta z
\rightarrow 0$. Hence
\begin{equation}\label{Eq:E_nei}
 E^i_{ne}(z,t) = \frac{4\pi}{3}{\wp}^i,
 \quad \text{with} \quad i=P,C.
\end{equation}

\subsection{Macroscopic Field in intermediate and far-off regions}
We estimate the contributions from the remaining fields replacing the summation
by an integration in Eq.~(\ref{Eq:MicroscopicField_L}) (see
Ref.~\cite{Bowden:1993})
\begin{equation}\label{Eq:MacroscopicField_M1}
\mathbf{E}_M^{i}(\mathbf{r},t) = \int d^3r'U^i(|\mathbf{r} - \mathbf{r'}|)
\exp[-i\mathbf{k}^i\cdot(\mathbf{r} - \mathbf{r'})]\,
\rho^i (\mathbf{r'}, t - |\mathbf{r} - \mathbf{r'}|/c).
\end{equation}
Notice that in this integral the contribution due to the nearby region is
excluded, i.e., the region inside the slab of length $\Delta z$ at $z$
is omitted. Since we are considering here the plane-wave approximation for the
field, then
\begin{equation}\label{Eq:rho_i}
\rho^i(\mathbf{r'},t - |\mathbf{r} - \mathbf{r'}|/c) =
\rho^i(z',t - |z-z'|/c),
\quad \text{with} \quad i=P,C.
\end{equation}
Substituting Eq.~(\ref{Eq:rho_i}) in Eq.~(\ref{Eq:MacroscopicField_M1}), and
assuming cylindrical geometry for the atomic sample, the field contribution can
be expressed as
\begin{equation}\label{Eq:MacroscopicField_M2}
\begin{split}
\mathbf{E}_M^{i}(z_l,t) = N \int_0^{z_l-d} \int_0^{2\pi} \int_0^{s_M}
s\,U^i (|z_l-z'|, s, \theta) e^{-i k^i(z_l - z')} \\
\times\,\rho^i (z', t - |z_l-z'|/c)\,ds\,d\theta\,dz',
\end{split}
\end{equation}
where
\begin{align}\label{Eq:}
U^i = (3/2) \zeta^i \bigg\{Q^i_1(k^i r)\sin^2 \theta
&+ Q^i_2(k^i r)\cos^2\theta \nonumber\\
&+ \frac{|z - z_l|^2}{r^2} [Q^i_1(k^i r) - Q^i_2(k^i r)]
\cos\theta \bigg\}.
\end{align}
Here, $r$ is as defined after Eq.~(\ref{Eq:H_i}); $Q_1^i$ and $Q_2^i$ are given
in Eq.~(\ref{Eq:Q_i}), assuming $r_{lm} \rightarrow r$. After integrating over
the variables $\theta$ and $s$, and using the parameter $r_M = (k^i)^2 s_M^2 +
(k^i)^2|z-z_l|^2$ with $i=P,C$, we get
\begin{align}\label{Eq:MacroscopicField_M3}
\begin{split}
\mathbf{E}_M^{i}(z_l,t) &= \frac{3\pi \zeta^i N}{(k^i)^2}
\int_0^{z_l-d}dz' \rho^i(z',t - |z_l-z'|/c) \\
&\qquad - \frac {3\pi \zeta^i n}{2(k^i)^2} \int_0^{z_l-d}dz'
e^{-ik^i(z_l-z')}\rho^i(z',t-|z_l-z'|/c) \\
&\qquad \times  e^{ir_{M}} \left[1+i(k^i)^2 \frac{|z'-z_l|^2}{r_M^3}\right],
\end{split}
\end{align}
In the limiting condition $k\,s_M \gg 1$, the second term in
Eq.~(\ref{Eq:MacroscopicField_M3}) is negligible giving
\begin{equation}\label{Eq:MacroscopicField_M4}
\mathbf{E}_M^{i}(z_l,t)=\frac {3\pi \zeta^i N}{(k^i)^2} \int_0^{z_l-d}dz'
\rho^i(z',t - |z_l-z'|/c),
\quad \text{with} \quad i = P,C.
\end{equation}
If we just want to estimate the contribution from the intermediate region, then
we need the condition $\Delta z \rightarrow 0$ and $z_l=d$ in
Eq.~(\ref{Eq:MacroscopicField_M4}). Using this condition, we get the expression
which is equal to the second term of Eq.~(\ref{Eq:Gi_2}). Also, under the
condition, $d \rightarrow 0$ and $z_l \rightarrow z$
Eq.~(\ref{Eq:MacroscopicField_M4}) provides an estimation of the macroscopic
field for both far and intermediate regions as
\begin{equation}\label{Eq:MacroscopicField_M5}
\mathbf{E}_M^{i}(z_l,t) = \frac{3\pi \zeta^n}{(k^i)^2}
\int_0^{z}dz' \rho^i(z',t - |z_l-z'|/c),
\quad \text{with} \quad i = P,C.
\end{equation}
Clearly, the contribution to macroscopic field comes from the retarded dipoles,
satisfying causality conditions.

We obtain Maxwell's equations in SVEA by substituting the proper partial
derivatives with time and position coordinates in
Eq.~(\ref{Eq:MacroscopicField_M5}), leading to
\begin{equation}\label{Eq:PartialE_Mi}
\frac{\partial E_M^i}{\partial t} + c\,\frac{\partial E_M^i}{\partial z} =
\frac{3\pi \zeta^i c \mu_i}{(k^i)^2}\,\rho^i.
\end{equation}
Here, $\rho^P = \rho_{12}$ and $\rho^C = \rho_{23}$ are off-diagonal density
matrix elements for probe and coupling transitions, respectively. Now, we can
combine all fields including Eq.~(\ref{Eq:E_nei}) to get the total field seen by
the atom at a general location $z$ as
\begin{equation}\label{Eq:MicroscopicField_L2}
E_L^i(z,t) = E_{ex}^i(z,t) + E_M^i(z,t) + \frac{4\pi}{3}\wp^i(z,t),
\quad \text{with} \quad i = P,C
\end{equation}
in which the evolution of the macroscopic field is governed by Maxwell's
equations. The expression obtained in Eq.~(\ref{Eq:MicroscopicField_L2}) is
incidentally identical to that of Lorentz-Lorentz corrections. However,
the physical reasons behind these two expressions are quite different. We
redefine the slowly varying envelopes in Eq.~(\ref{Eq:Liouville}) in view of
Eqs.~(\ref{Eq:Gi_2}) and~(\ref{Eq:MicroscopicField_L2}), where contributions of
the external driving field, macroscopic and atom's self-field have been taken
care off and thus we rewrite Eq.~(\ref{Eq:Liouville}) as
\begin{align}\label{Eq:DressedStates}
\begin{split}
  \dot{\rho}_{22} - \dot{\rho}_{11} =&
    - (\gamma_{23} + 2\gamma_{21})\rho_{22}
    + 2i\mu_{12}(\varepsilon_L^P)^*\rho_{21}
    - 2i\mu_{12}(\varepsilon_L^P)\rho_{12} \\
  & + i\mu_{23}(\varepsilon_L^C)^*\rho_{23}
    - i\mu_{23}(\varepsilon_L^C)\rho_{32}
    - \gamma_{31}(\rho_{33}-\rho_{11})
    - \gamma_{21}^D|\rho_{21}|^2 \\
  \dot{\rho}_{22} - \dot{\rho}_{33} =&
    - (2\gamma_{23} + \gamma_{21})\rho_{22}
    + i\mu_{12}(\varepsilon_L^P)^*\rho_{21}
    - i\mu_{12}(\varepsilon_L^P)\rho_{12} \\
  & + 2i\mu_{23}(\varepsilon_L^C)^*\rho_{23}
    - 2i\mu_{23}(\varepsilon_L^C)\rho_{32}
    - \gamma_{31}(\rho_{11}-\rho_{33})
    - \gamma_{23}^D|\rho_{23}|^2 \\
  \dot{\rho}_{23} =&
    - i[\Delta_C - \epsilon_c(\rho_{22}-\rho_{33})]\rho_{23}
    - [\gamma - (\gamma_{23}^D/2)(\rho_{22} - \rho_{33})]\rho_{23} \\
  & + i\mu_{23}(\varepsilon_L^C)(\rho_{22}-\rho_{33})
    - i\mu_{12}(\varepsilon_L^P)\rho_{13} \\
  \dot{\rho}_{21} =&
    - i[\Delta_P - \epsilon_p(\rho_{22}-\rho_{11})]\rho_{21}
    - [\gamma - (\gamma_{21}^D/2)(\rho_{22} - \rho_{11})]\rho_{21} \\
  & + i\mu_{12}(\varepsilon_L^P)(\rho_{22}-\rho_{11})
    - i\mu_{23}(\varepsilon_L^C)\rho_{31} \\
  \dot{\rho}_{31} =&
    - [\gamma_{31} + i(\Delta_P - \Delta_C)]\rho_{31}
    -i[\epsilon_p (\rho_{22} - \rho_{11})
    -  \epsilon_c (\rho_{22} - \rho_{33})]\rho_{31} \\
  & - i\mu_{23}(\varepsilon_L^C)^*\rho_{21}
    + i\mu_{12}(\varepsilon_L^P)\rho_{32}
\end{split}
\end{align}

\section{\label{sec:analytical}Analytical results under perturbative
approximation}
Next, we find out the steady-state solution of the above density matrix to get
expressions of linear and nonlinear susceptibilities of probe transition. We
follow the iterative approach for this purpose and represent a general density
matrix element as
$\rho_{ij}=\rho_{ij}^{(0)}+\rho_{ij}^{(1)}+\rho_{ij}^{(2)}+...$, such that the
higher order component is calculated with the help of lower order components. We
also assume that the coupling field is stronger than the probe field and
initially for the zeroth order all the atomic population is in the ground state,
e.g., $\rho_{11}^{(0)}\simeq 1$, $\rho_{22}^{(0)}\simeq 0$, and
$\rho_{33}^{(0)}\simeq 0$. Under all such approximation, it is possible to show
that
\begin{equation}\label{Eq:rho_32}
\rho_{32}^{(1)} =
\frac{i\Omega_P^*/2}{\gamma_\eff + i\Delta_c - i\epsilon_c(\rho_{22} -
\rho_{33})^{(0)}}\,\rho_{31}^{(1)}
\end{equation}
\begin{align}
\begin{split}
\rho_{31}^{(1)} =
&
- \frac{i\Omega_C^*/2}{\gamma_{31} - i(\Delta_p - \Delta_c)
+ i\left[\epsilon_p(\rho_{22} - \rho_{11})^{(0)}
- \epsilon_c(\rho_{22} - \rho_{33})^{(0)}\right]}\\
&
\times
\left\{1 + \frac{|\Omega_P|^2/4}{\gamma_{31} - i(\Delta_p - \Delta_c)
+ i\left[\epsilon_p(\rho_{22} - \rho_{11})^{(0)} - \epsilon_c(\rho_{22} -
\rho_{33})^{(0)}\right]}
\right. \\
&
\left.
\times \frac{1}{\gamma_\eff + i\Delta_c - i\epsilon_c(\rho_{22}
- \rho_{33})^{(0)}} \right\}^{-1} \rho_{21}^{(1)}
\end{split}
\end{align}

\begin{align}
\begin{split}
\rho_{21}^{(1)} =
&
\frac{(i\Omega_P)}{2}(\rho_{22}-\rho_{11})^{(0)}
\Bigg(\gamma_\eff - i\Delta_P + i\epsilon_P (\rho_{22}-\rho_{11})^{(0)}\\
&
+ \frac{|\Omega_C|^2/4}
{\gamma_{31} - i(\Delta_P - \Delta_C)
+ i[\epsilon_P (\rho_{22}-\rho_{11})^{(0)}
- \epsilon_C (\rho_{22}-\rho_{33})^{(0)}]}\\
&\qquad
\times
\left\{1 + \frac{|\Omega_P|^2/4}{\gamma_{31} - i(\Delta_p - \Delta_c)
+ i\left[\epsilon_p(\rho_{22} - \rho_{11})^{(0)} - \epsilon_c(\rho_{22} -
\rho_{33})^{(0)}\right]}
\right. \\
&\qquad\qquad
\left.
\times \frac{1}{\gamma_\eff + i\Delta_c - i\epsilon_c(\rho_{22}
- \rho_{33})^{(0)}} \right\}^{-1}
\Bigg)^{-1}\\
=&
\frac{(i\Omega_P/2)(\rho_{22}-\rho_{11})^{(0)}}{F}\\
\end{split}
\end{align}
where we have defined $\Omega_P$ and $\Omega_C$ after Eq.~(\ref{Eq:Liouville})
and
\begin{equation}
\gamma_\eff = \gamma -\frac{\gamma^D_{1}}{2}(\rho_{22} - \rho_{11})^{(0)}
\end{equation}
with
\begin{align}\label{Eq:F}
\begin{split}
F &=
\gamma_\eff - i\Delta_P + i\epsilon_p(\rho_{22} - \rho_{11})^{(0)}\\
&\quad
+ \frac{|\Omega_C|^2/4}{\gamma_{31} - i(\Delta_p - \Delta_c)
+ i\left[\epsilon_p(\rho_{22} - \rho_{11})^{(0)}
- \epsilon_c(\rho_{22} - \rho_{33})^{(0)}\right]}\\
&\qquad
\times
\left\{1 + \frac{|\Omega_P|^2/4}{\gamma_{31} - i(\Delta_p - \Delta_c)
+ i\left[\epsilon_p(\rho_{22} - \rho_{11})^{(0)} - \epsilon_c(\rho_{22} -
\rho_{33})^{(0)}\right]}
\right. \\
&\qquad\qquad
\left.
\times \frac{1}{\gamma_\eff + i\Delta_c - i\epsilon_c(\rho_{22}
- \rho_{33})^{(0)}} \right\}^{-1}
\end{split}
\end{align}
We obtain the third order component in $\rho _{21}$ (i.e., $\rho _{21}^{(3)}$),
after calculating $\rho_{22} - \rho_{11}$ to the second order [i.e.,
$(\rho_{22} - \rho_{11})^{(2)}$]. Thus
\begin{align}\label{Eq:rho_22-rho_21}
(\rho_{22} - \rho_{11})^{(2)}
&=
\frac{2}{2\gamma_\eff + \gamma_{21}}
\left[ i\Omega_P^* \rho_{21}^{(1)} -  i\Omega_P\rho_{12}^{(1)} \right]
+ \frac{2}{2\gamma_\eff + \gamma_{21}}, \nonumber\\
&=
\frac{2}{2\gamma_\eff+\gamma_{21}} \left(\frac{1}{F} + \frac{1}{F^*}\right)
\left[-|\Omega_P|^2(\rho_{22} - \rho_{11})^{(0)}\right] +
\frac{2\gamma_{31}}{2\gamma_\eff + \gamma_{21}}
\end{align}
\begin{align}
\begin{split}
\rho_{21}^{(3)} =
&
\frac{i\Omega_P}{2}
\left\{
\frac{1}{\gamma_\eff + \gamma_{21}}
\left(\frac{1}{F} + \frac{1}{F^*}\right)
\left[-|\Omega_P|^2(\rho_{22} - \rho_{11})^{(0)}\right] +
\frac{2\gamma_{31}}{2\gamma_\eff + \gamma_{21}}
\right\}\\
&
\times
\Bigg(
\gamma_\eff - i\Delta_P + \frac{i\epsilon_P}{2\gamma_\eff + \gamma_{21}}
\left\{
\left(\frac{1}{F} + \frac{1}{F^*}\right)
\left[-|\Omega_P|^2(\rho_{22} - \rho_{11})^{(0)}
\right]
+ 2\gamma_{31}
\right\}\\
&
+ \frac{|\Omega_C|^2/4}
{\gamma_{31} - i(\Delta_P - \Delta_C)
+ \frac{i\epsilon_P}{2\gamma_{eff} + \gamma_{21}}
\left\{\left(\frac{1}{F} + \frac{1}{F^*}\right)
\left[ -|\Omega_P|^2 (\rho_{22} -\rho_{11})^{(0)} \right]
+ 2\gamma_{31}
\right\}}
\Bigg)^{-1}
\end{split}
\end{align}
In getting these expressions, quantities like $\rho_{33}^{(2)}$ and so are
disregarded. Then, and up to third order of accuracy, the expression of
$\rho_{21}$ reads as
\begin{align}\label{Eq:rho_21_approx}
\begin{split}
\rho_{21} &\simeq \rho_{21}^{(1)} + \rho_{21}^{(3)} \\
&=
i\Omega_P(\rho_{22} - \rho_{11})^{(0)}
\Bigg\{
\frac{1}{F} +
\left[
\frac{-|\Omega_P|^2}{\gamma_\eff + \gamma_{21}}
\left(\frac{1}{F} + \frac{1}{F^*}\right)
+ \frac{2\gamma_{31}}{2\gamma_\eff + \gamma_{21}}
\right]\\
& \quad
\times
\bigg(
\gamma_\eff - i\Delta_P + \frac{i\epsilon_P}{2\gamma_\eff + \gamma_{21}}
\left\{
\left(\frac{1}{F} + \frac{1}{F^*}\right)
\left[-|\Omega_P|^2(\rho_{22} - \rho_{11})^{(0)}
\right]
+ 2\gamma_{31}
\right\}\\
& \quad
+ \frac{|\Omega_C|^2/4}
{\gamma_{31} - i(\Delta_P - \Delta_C)
+ \frac{i\epsilon_P}{2\gamma_{eff} + \gamma_{21}}
\left\{\left(\frac{1}{F} + \frac{1}{F^*}\right)
\left[ -|\Omega_P|^2 (\rho_{22} -\rho_{11})^{(0)} \right]
+ 2\gamma_{31}
\right\}}
\bigg)^{-1}
\Bigg\}
\end{split}
\end{align}
The macroscopic polarization of the atomic medium on the probe transition can
be expressed as
\begin{equation}\label{Eq:Macroscopic_P1}
P(M)_P = n \left(\rho_{12} \mu_{12} e^{-i\omega_P t}
+ \rho_{12}^* \mu_{12}^* e^{i\omega_P t}\right).
\end{equation}
Alternatively, the macroscopic polarization can also be written as
\begin{equation}\label{Eq:Macroscopic_P2}
P(M)_P = (1/2) \epsilon_0 \left(E_P \chi e^{-i\omega_P t}
+ E_{P}^* \chi^* e^{i\omega_P t}\right),
\end{equation}
where $\chi$ is the electrical susceptibility of the medium on the probe
transition. With the help of Eqs.~(\ref{Eq:Macroscopic_P1})
and~(\ref{Eq:Macroscopic_P2}), we get
\begin{equation}\label{Eq:chi1}
\chi = \frac{2n\mu_{12}\rho_{12}}{\epsilon_0 E_P},
\end{equation}
such that
\begin{equation}\label{Eq:chi2}
\chi = \chi^{(1)} + 3|E_P|^2 \chi^{(3)}.
\end{equation}
Then, we can write expressions for the first- and third-order susceptibilities
as
\begin{align}\label{Eq:chi3}
\chi^{(1)} &=
i\,\frac{n |\mu_{12}|^2}{\epsilon_0 \hbar}
\left[\frac{(\rho_{22} - \rho_{11})^{(0)}}{F} +
\frac{2\gamma_{31}}{(2\gamma_\eff + \gamma_{21})K}\right], \\
\chi^{(3)} &=
-i\frac{ n |\mu_{12}|^4 (\rho_{22} - \rho_{11})^{(0)}}
{3\epsilon_0 \hbar^3  (2\gamma_\eff + \gamma_{21})K}
\left(\frac{1}{F} + \frac{1}{F^*}\right),
\end{align}
where
\begin{align}
\begin{split}
K &=
\Bigg(
\gamma_\eff -
i\left\{ \Delta_P - \frac{\epsilon_P}{2\gamma_\eff + \gamma_{21}}
\left[\left(\frac{1}{F} + \frac{1}{F^*}\right)
(-|\Omega_P|^2(\rho_{22} - \rho_{11})^{(0)})
+ 2\gamma_{31}\right] \right\}\\
&
+ \frac{|\Omega_C|^2/4}{\gamma_{31} - i(\Delta_P-\Delta_C)
+ \frac{i\epsilon_P}{2\gamma_\eff + \gamma_{21}}
\left[ \left(\frac{1}{F} + \frac{1}{F^*}\right)
\left(-|\Omega_P|^2 (\rho_{22} - \rho_{11})^{(0)}\right)
+ 2\gamma_{31} \right]}\Bigg)
\end{split}
\end{align}
The analytic closed form results under the weak probe excitation are given
above providing the effect of NDD interaction on first- and third-order
susceptibilities of probe transitions. However, we focus our attention on the
exact results given in the following section.

\section{\label{sec:numerical}Numerical results}
In this section, we will study the steady-state results obtained by numerical
integration of Eq.~(\ref{Eq:DressedStates}) for the dynamical evolution of
density matrix elements, EIT and dispersive properties  of the $\lambda$-type
three-level system with NDD interaction included, under different parametric
conditions.

\subsection{Coherent Population Trapping (CPT) condition}
\begin{figure}
  \includegraphics[scale=0.6]{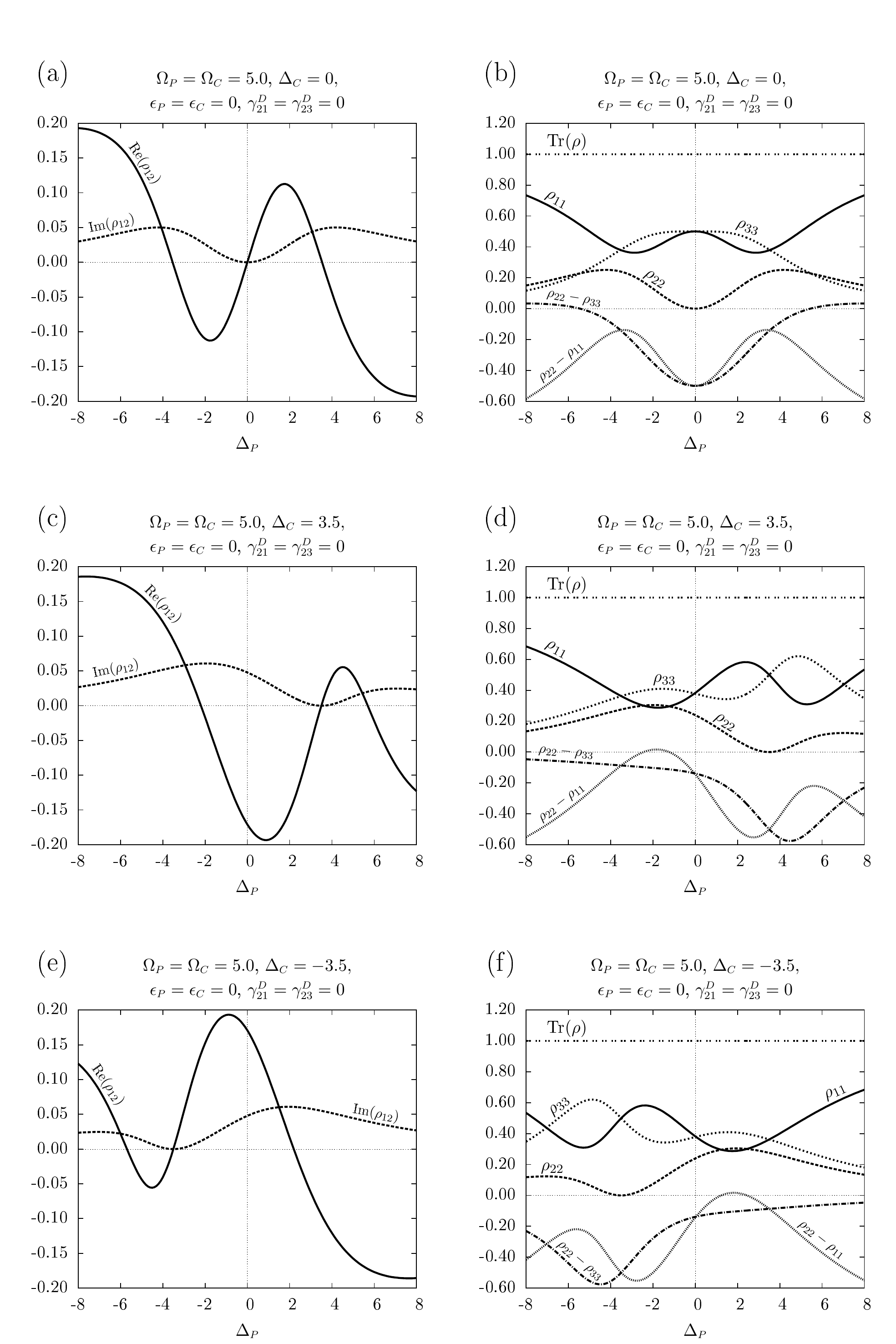}
  \caption{}
  \label{Fig:2}
\end{figure}

One can achieve CPT condition in the $\lambda$-type atomic system when the probe
Rabi frequency and coupling Rabi frequency are equal and thus satisfying
two-photon resonance condition. We keep $\Omega_P=\Omega_C=0.5$. All parameters
are measured in $\gamma$. Initially, $\rho_{11}=1$, $\rho_{22}=0$,
$\rho_{33}=0$. Fig.~\ref{Fig:2}(a) displays $\operatorname{Re}(\rho_{12})$ and
$\operatorname{Im}(\rho_{12})$ in the steady state as a function of probe
detuning $\Delta_P$ and keeping $\Delta_C=0$. The linear susceptibility of probe
transition is proportional to $\rho_{12}$, with dispersion proportional to
$\operatorname{Re}(\rho_{12})$ and absorption proportional to
$\operatorname{Im}(\rho_{12})$. Fig.~\ref{Fig:2}(a) shows typical dispersion
(solid line), absorption (dash line) curve under CPT condition. The peaks in the
absorption curve is due to the dynamic Stark-splitting of the upper level.
Fig.~\ref{Fig:2}(b) displays the plot of steady-state values of $\rho_{11}$,
$\rho_{22}$, $\rho_{33}$, $\rho_{22}-\rho_{11}$, $\rho_{22}-\rho_{33}$ and
$\Tr(\rho)$ as a function of $\Delta_P$, keeping all other parameters same as in
Fig.~\ref{Fig:2}(a). The CPT can be clearly observed in the region around
$\Delta_P=\Delta_C=0$, where two-photon resonance condition is satisfied, and
$\rho_{22}=0$ where $\rho_{11}=\rho_{33}=0.5$. The entire population is trapped
in two lower levels and the upper level is empty. The quantities
$\rho_{22}-\rho_{11}$ and $\rho_{22}-\rho_{33}$ show dips at $\Delta_P=0$ with
magnitude $-0.5$. Also, $\Tr{\rho}=1$ for the entire range of $\Delta_P$. The
effect of finite positive detuning of coupling field can be seen in
Fig.~\ref{Fig:2}(c), where $\Delta_C=3.5$ has been kept but keeping all other
parameters same as in Fig.~\ref{Fig:2}(a). The dip of absorption (dash line) and
peak of dispersion move toward $\Delta_P=3.5$.  Both dispersion and absorption
curves show their magnitudes zero at $\Delta_P=3.5$. In Fig.~\ref{Fig:2}(d) the
effect of  $\Delta_C=3.5$ can be seen on $\rho_{11}$, $\rho_{22}$, $\rho_{33}$,
$\rho_{22}-\rho_{11}$ and $\rho_{22}-\rho_{33}$, while keeping other parameters
same as in Fig.~\ref{Fig:2}(b). Here we find that CPT can be observed near
$\Delta_P=3.5$, where $\rho_{22}=0$ and $\rho_{11}=\rho_{33}=0.5$. The dips (or
peaks) of quantities $\rho_{22}- \rho_{11}$ and $\rho_{22}-\rho_{33}$ move in
opposite directions with respect to the point $\Delta_P=3.5$.The values of these
quantities at $\Delta_P=3.5$ is about $-0.5$. The $\Tr(\rho)=1$ for all values
of $\Delta_P$. Figures~\ref{Fig:2}(e) and~\ref{Fig:2}(f) are same
as~\ref{Fig:2}(c) and~\ref{Fig:2}(d) but this time $\Delta_C=-3.5$. The profile
of dispersion (solid line) curve is upside down and left right inverted when
compared to~\ref{Fig:2}(c). The absorption curve (dash line) is left right
inverted when compared to~\ref{Fig:2}(c). Both curves have zero magnitude at
$\Delta_P=-3.5$. Figure~\ref{Fig:2}(f) is left right inverted from
Fig.~\ref{Fig:2}(d). This means CPT is observable when $\Delta_P=-3.5$, i.e.,
$\rho_{22}=0$ and $\rho_{11}=\rho_{33}=0.5$ around $\Delta_P=-3.5$ (satisfying
two-photon resonance condition) along with $\rho_{22}-\rho_{11} =
\rho_{22}-\rho_{33} =-0.5$. Note $\Tr(\rho)=1$ throughout.

The effect of NDD interaction is shown in Figs.~\ref{Fig:3}--5. The parameters
selected for these figures are same as in Fig.~\ref{Fig:2} except $\epsilon_P =
\epsilon_C= 2.0$; $\gamma_{12}^D=\gamma_{23}^D=0$ for Figs.~\ref{Fig:3}(a,c) and
$\epsilon_P = \epsilon_C= 2.0$; $\gamma_{12}^D=\gamma_{23}^D=3.0$ for
Figs.~\ref{Fig:3}(b,d). Neglecting $\gamma_{12}^D$ and $\gamma_{23}^D$ is
justified under the limiting condition mentioned after Eq.~(\ref{Eq:Gi_2}). In
Fig.~\ref{Fig:3}(a) two curves represent dispersion (solid line) and absorption
(dash line) for the probe transition. When this figure is compared with
Fig.~\ref{Fig:2}(a) (where the NDD interaction is zero), there is some
noticeable change in absorption i.e., the curve getting flat but the peak in the
dispersion curve on the right-hand side is decreased due to non-zero real part
of NDD interaction. Similarly, for the diagonal matrix elements of $\rho$ when
compared with Fig.~\ref{Fig:2}(b), there is not much change in $\rho_{22}$ due
to the real part of NDD interaction but there is asymmetry introduced in
$\rho_{11}$ and $\rho_{22}$ due to that. The CPT is observable at $\Delta_P=0$
($=\Delta_C$) and at that point $\rho_{11}=\rho_{22}=0.5$. The peaks of both
$\rho_{11}$ and $\rho_{33}$ not occurring at $\Delta_P=0$. For $\rho_{11}$ it
has moved in negative $\Delta_P$ region. For $\rho_{33}$ there are two peaks of
unequal magnitudes. Also, dips in $\rho_{22}-\rho_{11}$, $\rho_{22}-\rho_{33}$
move in to the opposite directions with reference to $\Delta_P=0$ (as compared
to Fig.~\ref{Fig:2}(b)). Effects of non-zero NDD interaction when its both real
and imaginary parts are non-zero is shown in Fig.~\ref{Fig:3}(c) for dispersion
(solid line) and absorption (dash line). There are some changes in the widths of
absorption/dispersion spectra and magnitudes of curves change when compared with
Fig.~\ref{Fig:2}(a). This is due to the enhanced damping introduced by imaginary
part of the NDD interaction. The same is true for Fig.~\ref{Fig:3}(d) where some
changes are noticeable in $\rho_{11}$, $\rho_{22}$, $\rho_{33}$. The CPT is
observable at $\Delta_P=0$ ($=\Delta_C$), where $\rho_{11}=\rho_{22}=0.5$ and
asymmetry in $\rho_{11}$ and $\rho_{33}$  curves is still observable with a
slight change with respect to Fig.~\ref{Fig:3}(b). The widths of
$\rho_{22}-\rho_{11}$, $\rho_{22}-\rho_{33}$ curves also increasing in
Fig.~\ref{Fig:3}(d) in comparison to Fig.~\ref{Fig:3}(b).

Next, the effect of non-zero $\Delta_C$ along with NDD interaction is displayed
in Fig.~\ref{Fig:4}(a,b,c,d), where $\Delta_C=3.5$ is selected but all other
parameters are the same as in Fig.~\ref{Fig:3}. The dip in the absorption curve
(dash line) shows going down and up behavior at zero absorption
(Fig.~\ref{Fig:4}(a,c)), which is occurring near $\Delta_P=3.5$, when compared
with Fig.~\ref{Fig:2}(c). The dispersion curve also rises sharply when
approaching $\Delta_P=3.5$ (solid curve in Fig.~\ref{Fig:4}(a) when compared to
solid curve in Fig.~\ref{Fig:2}(c). This is due to the non-zero $\epsilon_P$ and
$\epsilon_C$. The inclusion of non-zero $\gamma_{12}^D$ and $\gamma_{23}^D$
causes the dispersion curve to go down in magnitude with increased width of
peaks (solid curve in Fig.~\ref{Fig:4}(c) when compared to solid curve in
Fig.~\ref{Fig:4}(a)). There are both qualitative and quantitative changes in
curves of diagonal matrix elements $\rho_{11}$, $\rho_{22}$, $\rho_{33}$ in the
presence of NDD interaction and $\Delta_C=3.5$ as shown in Fig.~\ref{Fig:4}(b).
The most noticeable behavior of these curves are near $\Delta_P=3.5$ where fall
of $\rho_{11}$, and rise of $\rho_{33}$ is quite steep and the dip in
$\rho_{22}$ is not smooth. This is all due to the combined effect of $\epsilon_P
= \epsilon_C= 2.0$ and $\Delta_C=3.5$. When $\gamma_{12}^D = \gamma_{23}^D=3.0$
is included in Fig.~\ref{Fig:4}(d) then magnitudes get diminished. Such behavior
are also seen in $\rho_{22}-\rho_{11}$, $\rho_{22}-\rho_{33}$ near
$\Delta_P=3.5$. In Figs.~\ref{Fig:4}(b,d) the CPT is occurring near
$\Delta_P=3.5$ such that $\rho_{22}=0$, $\rho_{11}= \rho_{33}=0.5$.

In Fig.~\ref{Fig:5}(a,b,c,d) the value of $\Delta_C=-3.5$ but all other
parameters are the same as in Fig.~\ref{Fig:4}(a,b,c,d). The dispersion curve
(solid line) when compared to dispersion curve in Fig.~\ref{Fig:4}(a) is now
up-side down inverted as well as left-right inverted. The zero dispersion is
located at $\Delta=-3.5$. The absorption curve (dash line) in
Fig.~\ref{Fig:5}(a) is left-right inverted in comparison to corresponding curve
in Fig.~\ref{Fig:4}(a). The zero absorption (dip) occurs at $\Delta=-3.5$. The
curve is quite smooth around the dip. In Fig.~\ref{Fig:5}(b) all the diagonal
matrix elements are left-right inverted in comparison to corresponding curves in
Fig.~\ref{Fig:4}(b). $\rho _{11}=\rho_{33}=0.5$ at $\Delta_P=-3.5$ and curves
are smoothly rising near that point. Also, both $\rho_{22}-\rho_{11}$ and
$\rho_{22}-\rho_{33}$ equal to $0.5$ at $\Delta_P=-3.5$ with smooth rise near
that point in these curves. When imaginary part of the NDD interaction is
included Fig.~\ref{Fig:5}(c,d), the magnitude of curves decrease at every point
that include peaks and dips when compared to Fig.~\ref{Fig:4}(c,d). This is due
to the increased damping caused by the $\gamma^D$(s).

\subsection{Electromagnetically Induced Transparency (EIT) condition}
The study of the EIT system under consideration is displayed in
Figs.~\ref{Fig:5}--\ref{Fig:9} with different parametric conditions. In all
these studies, the initial condition ($t=0$) for the density matrix elements has
been kept as $\rho_{11}(0)=1$, $\rho_{22}(0)=\rho_{33}(0)=0$, and
$\rho_{ij}(0)=0$ $i \neq j$. For Fig.~\ref{Fig:6}(a,b), other parameters are
$\Omega_P=0.1$, $\Omega_C=5.0$, $\epsilon_P=\epsilon_C=0$,
$\gamma_{21}^D=\gamma_{23}^D=0$ and all these parameters are measured with
respect to $\gamma$. The dispersion (solid line) and absorption (dash line)
curves are plotted in Fig.~\ref{Fig:6}(a). These curves are typical EIT
dispersion-absorption curves~\cite{Marangos:45,Fleischhauer:77}. The absorption
peaks are situated at $\Delta_P=\pm 2.5$. Both dispersion and absorption curves
are at zero magnitude when $\Delta_P=0$. Under EIT conditions, the population of
three levels in the steady state are $\rho_{11}=1$, $\rho_{22}=0$,
$\rho_{33}=0$, $\rho_{22}-\rho_{11}=-1$, $\rho_{22}-\rho_{33}=0$ as shown in
Fig.~\ref{Fig:6}(b). Figure~\ref{Fig:6}(a,b) serve as reference for comparison
with Fig.~\ref{Fig:6}(c,d,e,f) when $\Delta_C \neq 0$. Effect of $\Delta_C=3.5$
is shown in Fig.~\ref{Fig:6}(c,d) in which all other parameters are kept same as
in Fig.~\ref{Fig:6}(a,b) but $\Delta_C$ is different from zero. The non-zero
$\Delta_C$ causes asymmetry in absorption/dispersion spectra. Both locations and
separation of peaks change. Peaks of absorption are now located at $\Delta_P=$
4.8 and 1.3 consistent with the analysis given
in~\cite{Marangos:45,Fleischhauer:77}. The zero magnitude of both absorption
/dispersion now moves at $\Delta_C=3.5$. The non-zero $\Delta_C$ introduces
profile change of both absorption/dispersion curves. The behavior of diagonal
matrix elements of $\rho$ and their difference (Fig.~\ref{Fig:6}(d) is the same
as Fig.~\ref{Fig:6}(b). When the sign of $\Delta_C$ is reversed, i.e.,
$\Delta_C=-3.5$, then in Fig.~\ref{Fig:6}(e) it can be seen that absorption
spectrum is left-right reversed when compared with Fig.~\ref{Fig:6}(c). However
the dispersion spectrum in Fig.~\ref{Fig:6}(e) is not only left-right reversed
but also upside down when compared with Fig.~\ref{Fig:6}(c). Note that
Fig.~\ref{Fig:6}(f) remain same as Fig.~\ref{Fig:6}(b) for the diagonal
elements.

In Fig.~\ref{Fig:7}(a,b,c,d) the effect of NDD interaction has been displayed
under the EIT conditions. The parameters in Fig.~\ref{Fig:7}(a,b) are the same
as in Fig.~\ref{Fig:6}(a,b) except $\epsilon_P=\epsilon_C=2.0$, i.e., the real
part of the NDD interaction is non-zero in Fig.~\ref{Fig:7}(a,b). The profiles
of absorption (dash-line) and dispersion (solid-line) curves changed
(Fig.~\ref{Fig:7}(a)) compared to Fig.~\ref{Fig:6}(a) and they look similar to
the one displayed in Fig.~\ref{Fig:6}(c). The only difference is in the
locations of the peaks. In the steady-state, under the EIT conditions
$\rho_{11}=1$, $\rho_{22}=0$, $\rho_{33}=0$, $\rho_{22}-\rho_{11}=-1$,
$\rho_{22}-\rho_{33}=0$ as shown in Fig.~\ref{Fig:7}(b). The coefficient
$\epsilon_P$ comes in the expression as $\epsilon_P (\rho_{22}-\rho_{11})$ and
survives in the steady-state but $\epsilon_C$ comes in the expression as
$\epsilon_C (\rho_{22}-\rho_{33})$ and vanishes in the steady-state. Hence only
$\epsilon_P$ is retained in the equations of density matrix elements and it
behaves like atomic detuning in the steady-state. Fig.~\ref{Fig:7}(b) behaves
like Fig.~\ref{Fig:6}(b) more or less meaning $\rho_{11}=1$, $\rho_{22}=0$,
$\rho_{33}=0$, $\rho_{22}-\rho_{11}=-1$, $\rho_{22}-\rho_{33}=0$ in the
steady-state. Inclusion of the imaginary part of the NDD interaction
$\gamma_{21}^D=\gamma_{23}^D=3.0$ reduces the peak heights but increases the
widths of absorption/dispersion curves in Fig.~\ref{Fig:7}(c) when compared to
Fig.~\ref{Fig:7}(a). Clearly, this is due to the enhancement of damping caused
by the imaginary part of the NDD interaction. The diagonal elements’ behavior
Fig.~\ref{Fig:7}(d) is the same as the one displayed in Fig.~\ref{Fig:7}(b).

In Fig.~\ref{Fig:8}(a,b) the real part of the NDD interaction as well as other
parameters are the same as in Fig.~\ref{Fig:7}(a,b), except $\Delta_C=-3.5$. The
effective detuning is the algebraic sum of $\Delta_C=3.5$ and $\epsilon_P$ under
steady-state in this situation. Hence the separation of right peaks for both
dispersion/absorption curves from $\Delta_P=0.0$ increases. Also, in
Fig.~\ref{Fig:8}(a) the relative separation of two peaks of
absorption/dispersion curves also enhances. However,  Fig.~\ref{Fig:8}(b) is
quite similar to Fig.~\ref{Fig:7}(b) for the diagonal elements. When imaginary
part of the NDD interaction is also included by keeping
$\gamma_{21}^D=\gamma_{23}^D=3.0$ (in Fig.~\ref{Fig:8}(b)) and comparison of
curves made with Fig.~\ref{Fig:8}(a), the locations of the peaks are unchanged
but their widths increase due to the increased damping. The behavior of the
diagonal elements (Fig.~\ref{Fig:8}(d)) is same as the one displayed in
Fig.~\ref{Fig:8}(b).

Next, the $\Delta_C$ is selected as $\Delta_C=-3.5$ in Fig.~\ref{Fig:9}(a,b,c,d)
but all other parameters are the same as in Fig.~\ref{Fig:8}(a,b,c,d). By
selecting such parameters, the two peaks in the absorption/dispersion curves
look identical in shapes and heights and they are symmetrically located around
$\Delta_P \approx -2$. The zero absorption and dispersion are also located
around the same value of $\Delta_P$. The non-zero $\epsilon_P$ and
$\Delta_C=-3.5$ combined algebraically and move curves toward
$\Delta_P\approx-2$. Fig.~\ref{Fig:9}(b) shows $\rho_{11}=1$, $\rho_{22}=0$,
$\rho_{33}=0$, $\rho_{22}-\rho_{11}=-1$, $\rho_{22}-\rho_{33}=0$, in the
steady-state for all values of $\Delta_P$. When imaginary part of the NDD is
also included (Fig.~\ref{Fig:9}(c)), the symmetry, locations, shapes of all
curves still maintained as in Fig.~\ref{Fig:9}(a) but the peak heights decrease
due to the enhanced damping provided by the imaginary part of the NDD. Note that
there is no change in Fig.~\ref{Fig:9}(d)  when compared with
Fig.~\ref{Fig:9}(b).

\section{\label{sec:conclusion}Conclusion}
An ensemble of three-level atoms in $\lambda$-type configuration of its levels
is considered here. The near dipole-dipole interaction (NDD) for both probe and
coupling transitions are included in the density matrix equation of the system.
The NDD interactions give rise to a complex coefficient, whose real part
provides the frequency shift/detuning and imaginary part introduces an
additional radiative damping. The system has been studied under the steady-state
condition for both coherent-population-trapping (CPT) and electromagnetically
induced transparency (EIT) conditions. In the CPT condition, the lower
two-levels are equally populated when $\Delta_P=\Delta_C=0$ but that changes
when $\Delta_P$ is non-zero for both $\Delta_C=0$ and $\Delta_C \neq 0$. There
are both qualitative and quantitative changes in absorption/dispersion spectra
of the probe transition when NDD interaction is non-zero and $\Delta_C \neq 0$.
In the EIT condition, the probe field is kept weak compared to the coupling
field, and hence in the steady-state, the entire population stayed in the lower
level of the probe transition ($\rho_{11} = 1$). The real part of the NDD
interaction provides atomic detuning-like behavior for the probe transition but
no effect for the coupling transition. The absorption/dispersion change both
qualitatively as well as quantitatively under different values of parameters
(specifically for the change of $\Delta_C$ and $\epsilon_P$). Inclusion of the
imaginary part of the NDD interaction enhances radiative damping and
consequently the increase in peak widths and reduction of their amplitudes.
These results provide control of absorption/dispersion properties of the
three-level system for CPT and EIT phenomena with NDD interactions (in dense
atomic medium) and atomic detuning.

\section*{Acknowledgments}
We thank G. Duree for his encouragement and support for this work.


\newpage
\section*{Figure Captions}
{\bf Figure 1:} Diagram of the three-level system in a
$\Lambda$-configuration of its levels, driven by probe and coupling lasers of
frequencies $\omega_P$ and $\omega_C$, respectively.

{\bf Figure 2:} Plots of density matrix elements as a function of the probe
detuning $\Delta_P$ without NDD interaction under parametric conditions
$\Omega_P =\Omega_C=5.0$ (CPT condition), $\epsilon_P=\epsilon_C=0$, and
$\gamma_{21}^D=\gamma_{23}^D=0$. All parameters are measured with respect to
$\gamma$. Plots (a), (c), and (e) are for the real and imaginary parts of
$\rho_{12}$, which are proportional to the real and imaginary parts of the first
order susceptibility of the probe transition when $\Delta_C=0, 3.5,$ and $-3.5$,
respectively. Plots (b), (d), and (f) show $\rho_{11}$, $\rho_{22}$,
$\rho_{33}$, $\rho_{22}-\rho_{11}$, $\rho_{22}-\rho_{33}$, and $\text{Tr}(\rho)$
when $\Delta_C=0, 3.5$, and $-3.5$, respectively.

{\bf Figure 3:} Plots of density matrix elements as a function of the probe
detuning $\Delta_P$ with NDD interaction under parametric conditions $\Omega_P
=\Omega_C=5.0$ (CPT condition), $\epsilon_P=\epsilon_C=2.0$, and $\Delta_C=0$.
All parameters are measured with respect to $\gamma$. Plots (a) and (c) are for
the real and imaginary parts of $\rho_{12}$, which are proportional to the real
and imaginary parts of the first order susceptibility of the probe transition
when $\gamma_{21}^D=\gamma_{23}^D=0$ and $3.0$, respectively. Plots (b) and (d)
show $\rho_{11}$, $\rho_{22}$, $\rho_{33}$, $\rho_{22}-\rho_{11}$,
$\rho_{22}-\rho_{33}$, and $\text{Tr}(\rho)$ when
$\gamma_{21}^D=\gamma_{23}^D=0$ and $3.0$, respectively.

\begin{figure}
  \includegraphics[scale=0.6]{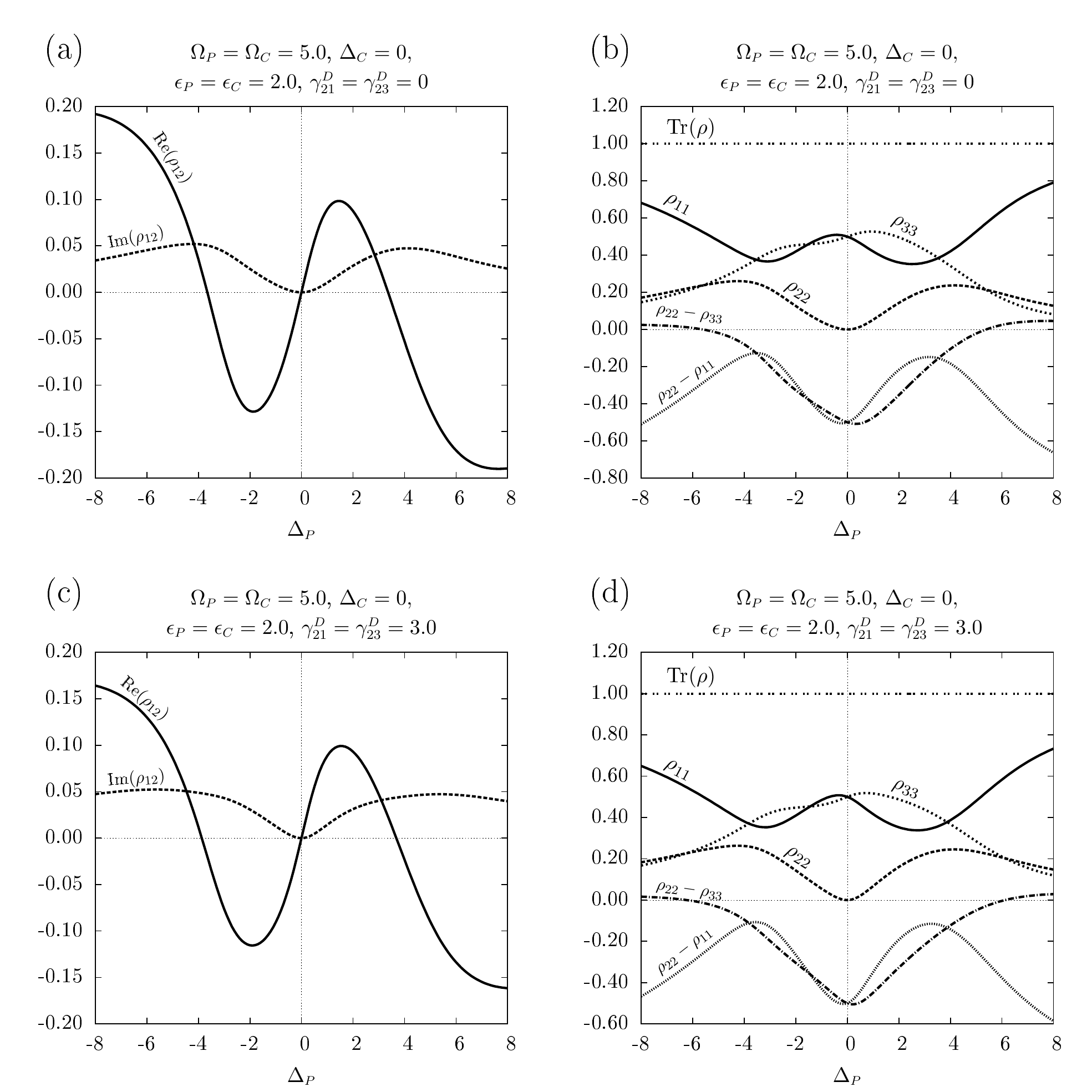}
  \caption{}
  \label{Fig:3}
\end{figure}

\begin{figure}
  \includegraphics[scale=0.6]{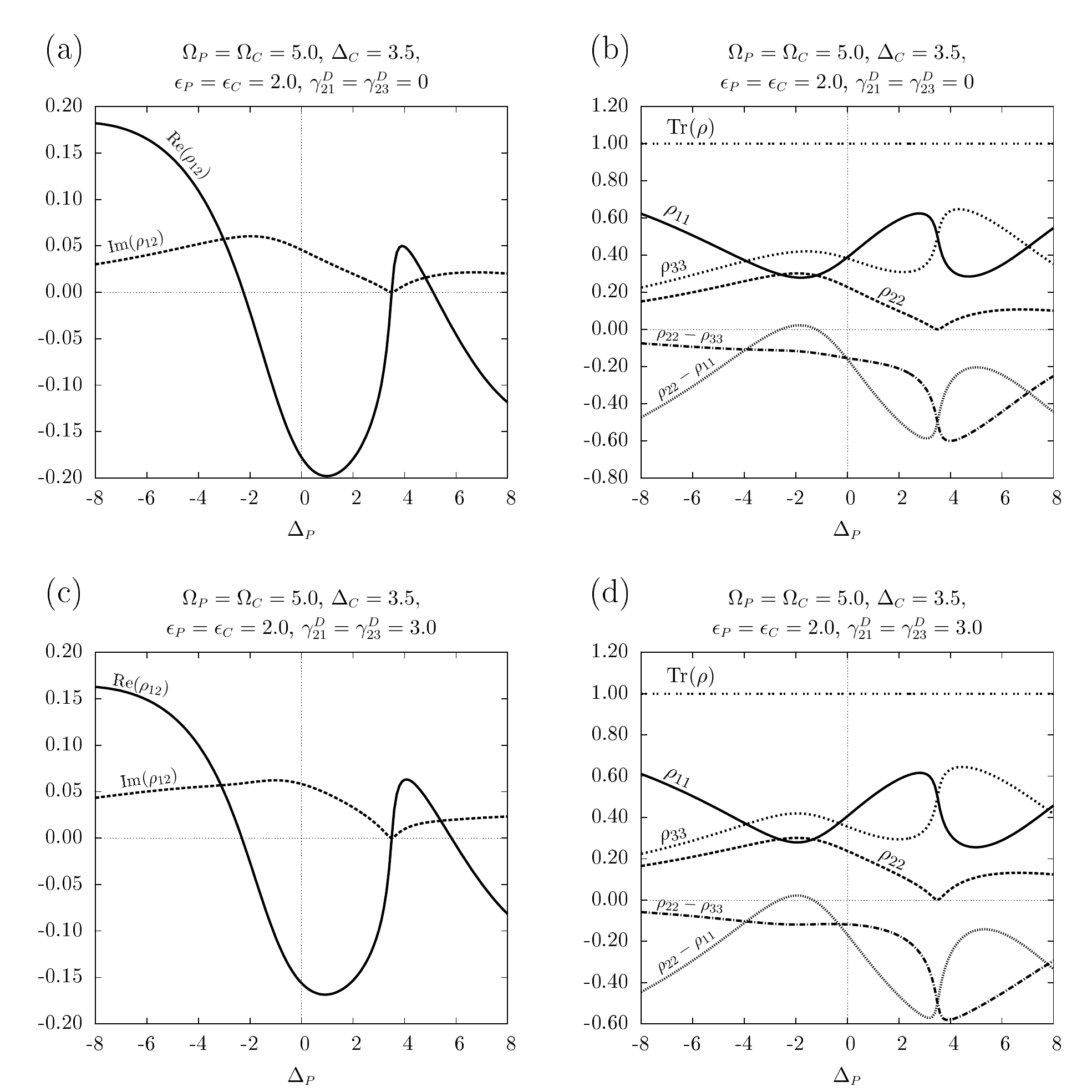}
  \caption{}
  \label{Fig:4}
\end{figure}

\begin{figure}
  \includegraphics[scale=0.6]{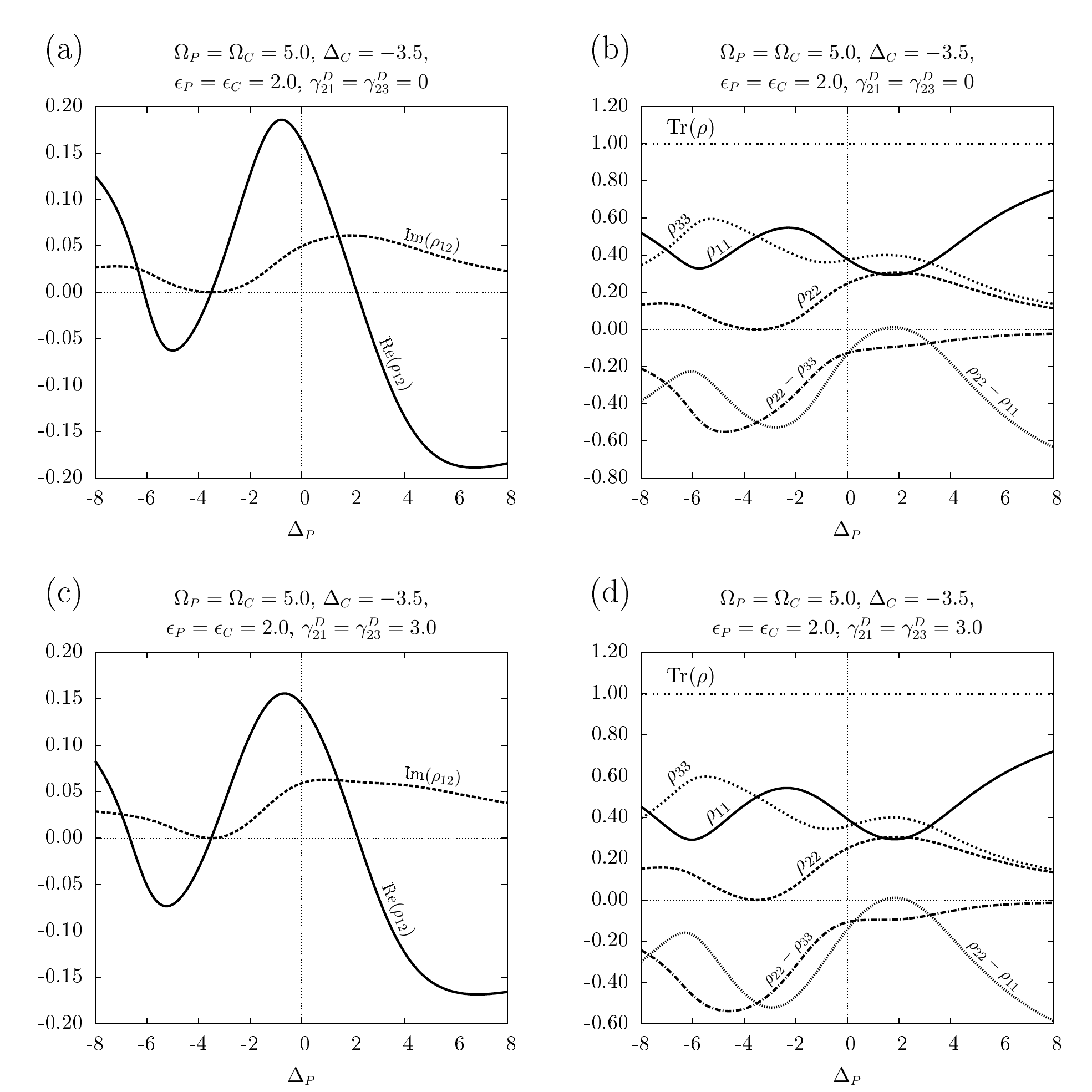}
  \caption{}
  \label{Fig:5}
\end{figure}

\begin{figure}
  \includegraphics[scale=0.6]{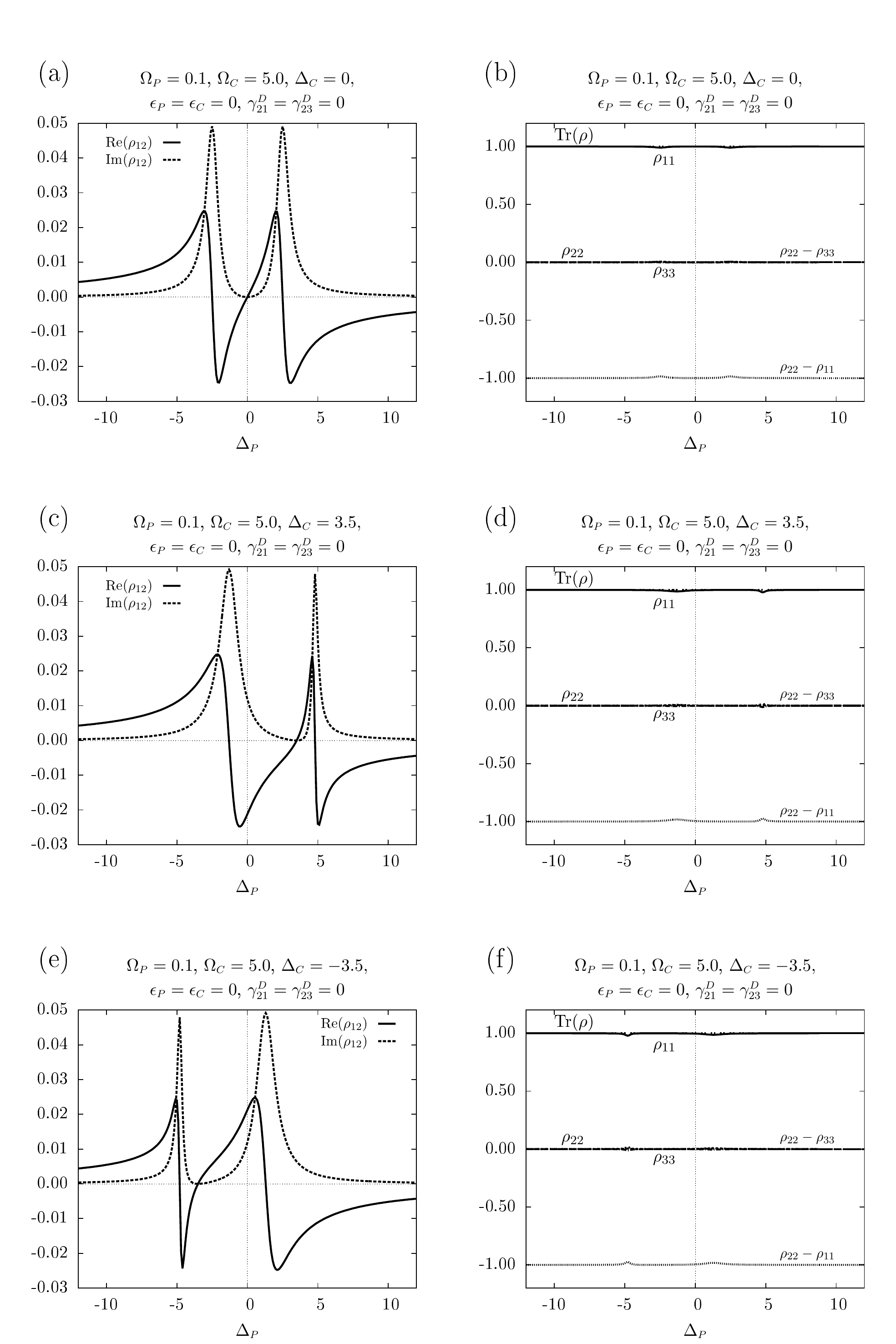}
  \caption{}
  \label{Fig:6}
\end{figure}

{\bf Figure 4:} Same as Fig.~\ref{Fig:3} but with $\Delta_C=3.5$.

{\bf Figure 5:} Same as Fig.~\ref{Fig:3} but with $\Delta_C=-3.5$.

{\bf Figure 6:} Plots of density matrix elements as a function of the probe
detuning $\Delta_P$ without NDD interaction under parametric conditions
$\Omega_P =0.1, \Omega_C=5.0$ (EIT condition), $\epsilon_P=\epsilon_C=0$, and
$\gamma_{21}^D=\gamma_{23}^D=0$. All parameters are measured with respect to
$\gamma$. Plots (a), (c), and (e) are for the real and imaginary parts of
$\rho_{12}$, which are proportional to the real and imaginary parts of the first
order susceptibility of the probe transition when $\Delta_C=0, 3.5,$ and $-3.5$,
respectively. Plots (b), (d), and (f) show $\rho_{11}$, $\rho_{22}$,
$\rho_{33}$, $\rho_{22}-\rho_{11}$, $\rho_{22}-\rho_{33}$, and $\text{Tr}(\rho)$
when $\Delta_C=0, 3.5$, and $-3.5$, respectively.

{\bf Figure 7:} Plots of density matrix elements as a function of the probe
detuning $\Delta_P$ with NDD interaction under parametric conditions $\Omega_P
=0.1, \Omega_C=5.0$ (EIT condition), $\epsilon_P=\epsilon_C=2.0$, and
$\Delta_C=0$. All parameters are measured with respect to $\gamma$. Plots (a)
and (c) are for the real and imaginary parts of $\rho_{12}$, which are
proportional to the real and imaginary parts of the first order susceptibility
of the probe transition when $\gamma_{21}^D=\gamma_{23}^D=0$ and $3.0$,
respectively. Plots (b) and (d) show $\rho_{11}$, $\rho_{22}$, $\rho_{33}$,
$\rho_{22}-\rho_{11}$, $\rho_{22}-\rho_{33}$, and $\text{Tr}(\rho)$ when
$\gamma_{21}^D=\gamma_{23}^D=0$ and $3.0$, respectively.

\begin{figure}
  \includegraphics[scale=0.6]{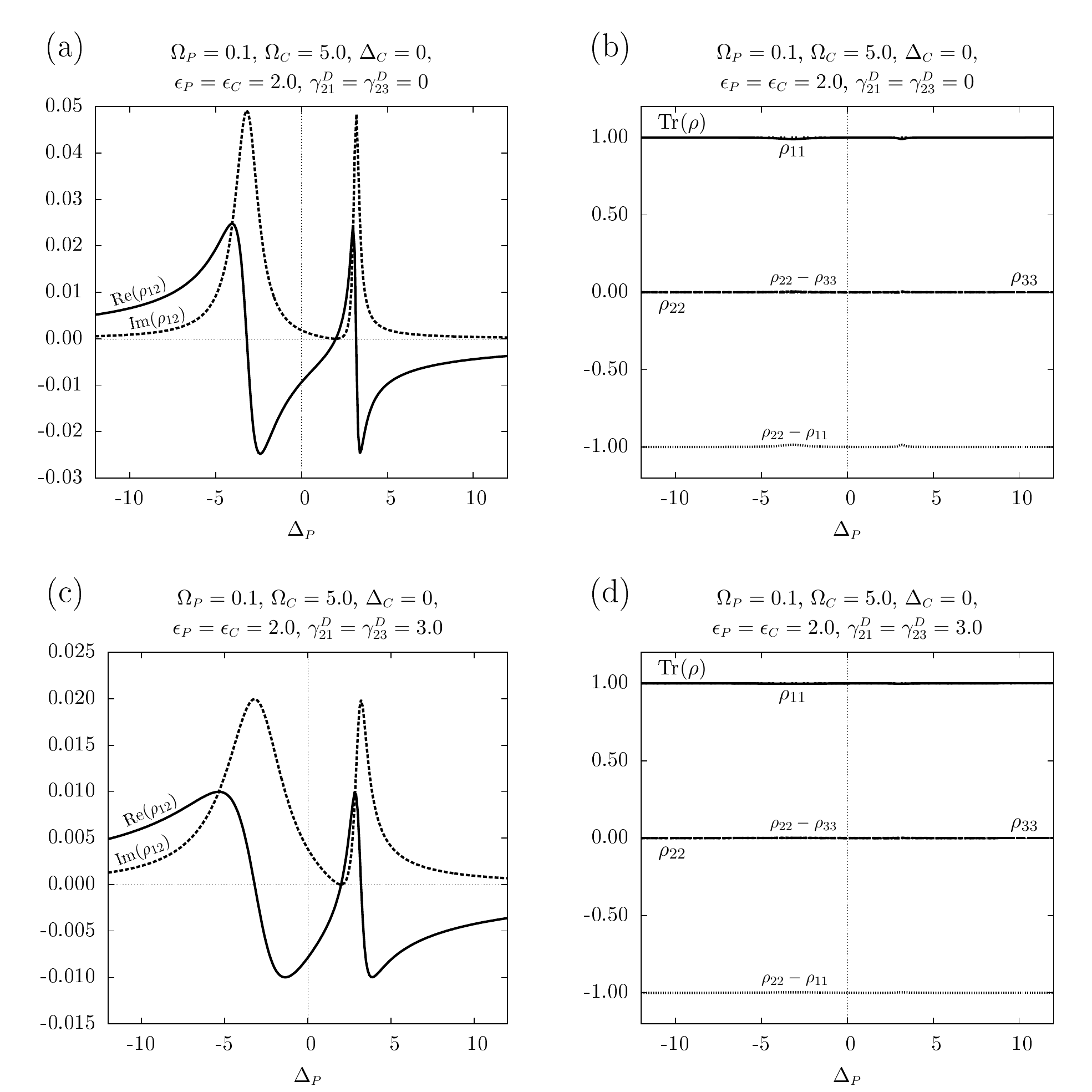}
  \caption{}
  \label{Fig:7}
\end{figure}

\begin{figure}
  \includegraphics[scale=0.6]{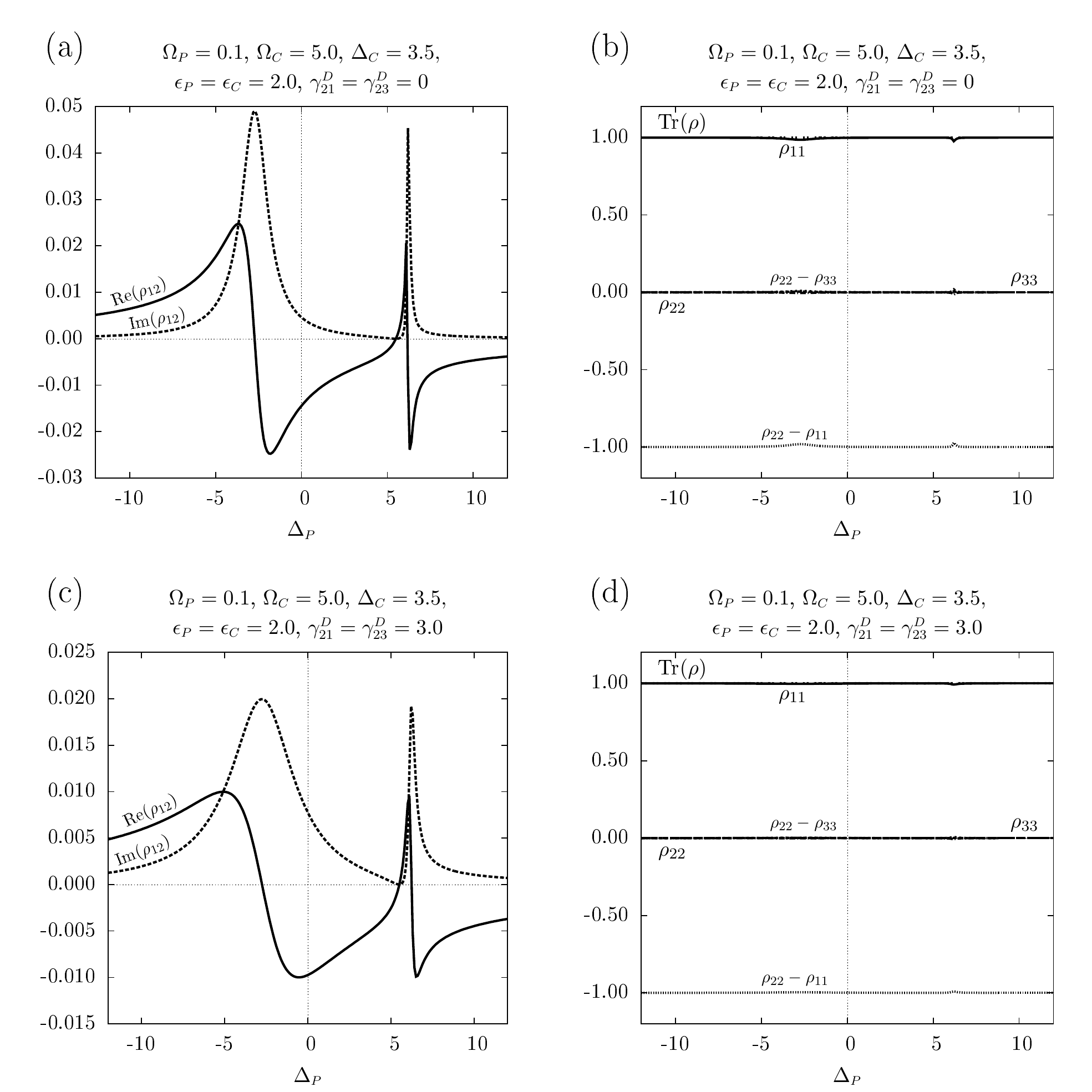}
  \caption{}
  \label{Fig:8}
\end{figure}

\begin{figure}
  \includegraphics[scale=0.6]{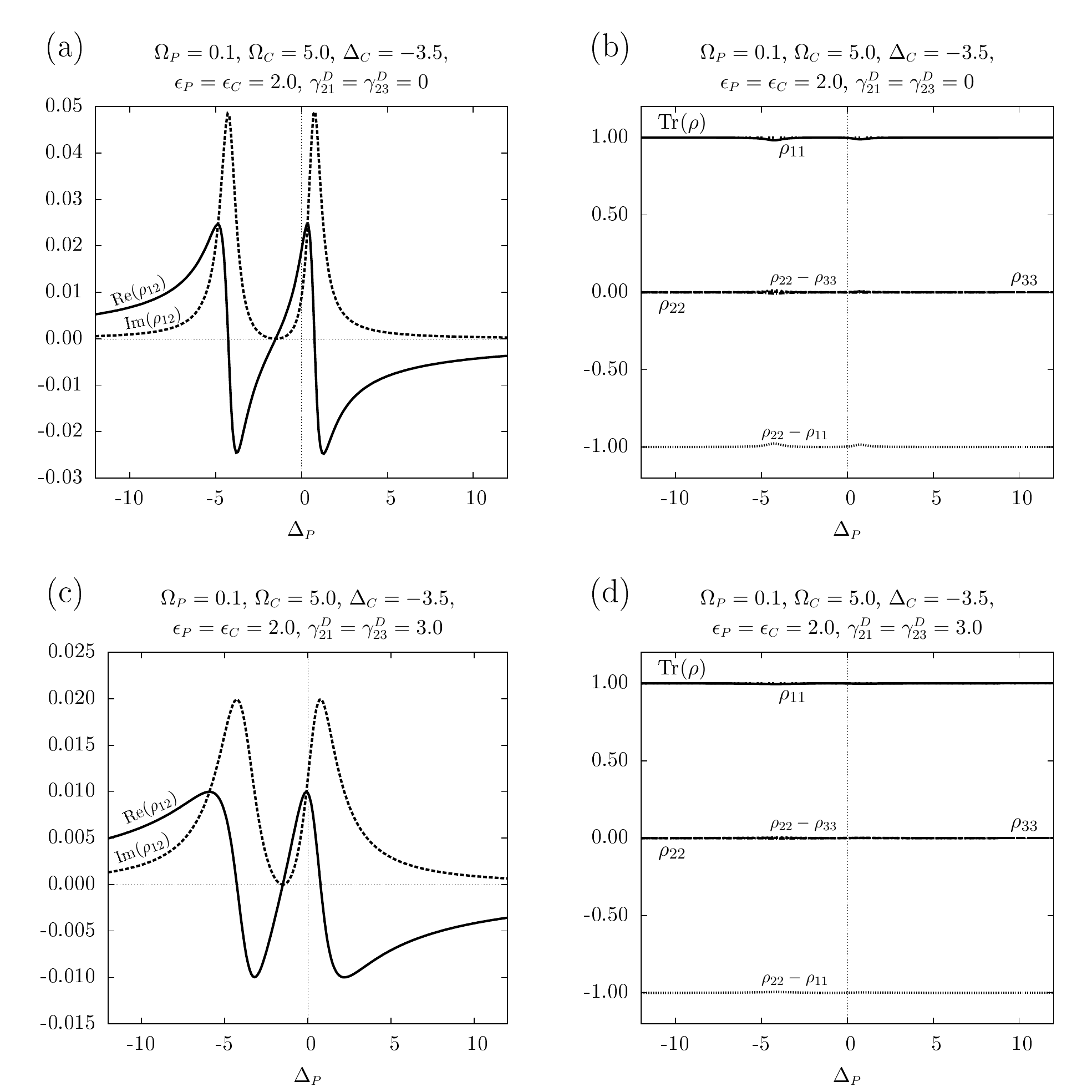}
  \caption{}
  \label{Fig:9}
\end{figure}

{\bf Figure 8:} Same as Fig.~\ref{Fig:7} but with $\Delta_C=3.5$.

{\bf Figure 9:} Same as Fig.~\ref{Fig:7} but with $\Delta_C=-3.5$.

\newpage

\clearpage

\clearpage

\clearpage

\clearpage

\clearpage

\clearpage

\clearpage

\clearpage

\end{document}